\documentclass[trackchanges,twocolumn,twocolappendix]{aastex701}

\usepackage{amsmath}    
\usepackage{graphicx}   
\usepackage{hyperref}   
\usepackage{gensymb}    
\usepackage{array}      
\usepackage{booktabs}   
\usepackage{multirow}
\usepackage{float}
\usepackage{footmisc}

\newcommand{\teff}{$T_{\rm{eff}}$}
\newcommand{\numax}{$\nu_{\rm{max}}$}
\newcommand{\dnu}{$\Delta\nu$}

\begin{document}

\title{Detecting Solar-Like Oscillations in the Highest Mass TESS Giants}

\author[0009-0006-8874-3846]{Noah J. Downing}
\affiliation{Department of Astronomy, Yale University, New Haven, CT 06520, USA}
\affiliation{Department of Astronomy, The Ohio State University, Columbus, OH 43210, USA}
\email[show]{noah.downing@yale.edu}

\author[0000-0003-0929-6541]{Madeline Howell}
\affiliation{Department of Astronomy, The Ohio State University, Columbus, OH 43210, USA}
\affiliation{Center for Cosmology and Astroparticle Physics (CCAPP), The Ohio State University, 191 West Woodruff Avenue, Columbus, OH 43210, USA}
\email{howell.753@osu.edu}

\author[0000-0002-7549-7766]{Marc H. Pinsonneault}
\affiliation{Department of Astronomy, The Ohio State University, Columbus, OH 43210, USA}
\affiliation{Center for Cosmology and Astroparticle Physics (CCAPP), The Ohio State University, 191 West Woodruff Avenue, Columbus, OH 43210, USA}
\email{pinsonneault.1@osu.edu}

\author[0000-0002-8854-3776]{Rafael A. Garc\'ia}
\affiliation{Université Paris-Saclay, Université Paris Cité, CEA, CNRS, AIM, 91191, Gif-sur-Yvette, France}
\email{rafael.garcia@cea.fr}

\author[0000-0002-6812-4443]{Dinil B. Palakkatharappil}
\affiliation{Université Paris-Saclay, Université Paris Cité, CEA, CNRS, AIM, 91191, Gif-sur-Yvette, France}
\email{dinil-bose.palakkatharappil@universite-paris-saclay.fr}

\author[0009-0008-1032-5067]{Lina Borg}
\affiliation{Université Paris-Saclay, Université Paris Cité, CEA, CNRS, AIM, 91191, Gif-sur-Yvette, France}
\email{lina.borg@cea.fr}

\author[0000-0002-0129-0316]{Savita Mathur}
\affiliation{Instituto de Astrofísica de Canarias (IAC), E-38205 La Laguna, Tenerife, Spain}
\affiliation{Departamento de Astrofísica,Universidad de La Laguna (ULL), E-38206 La Laguna, Tenerife, Spain}
\email{smathur@iac.es}

\begin{abstract}
Red-giant asteroseismology yields precise stellar parameters, making it a powerful tool for studying stellar structure and evolution, as demonstrated by the \textit{Kepler} mission. However, due to \textit{Kepler’s} limited field of view, it primarily sampled the more populous low-mass red giants found outside of the Galactic plane, leading to limited detections of red giants above $\rm 3\ M_{\odot}$. Here we use the all-sky TESS data to isolate 227 intermediate-mass candidates from large catalogs with a pre-selection based on photometric and spectroscopic data. We optimize TESS light curves using a boutique light curve detrending method with custom apertures. Compared to the MIT Quick Look Pipeline, this yields a 12\% average increase in the power-to-background ratio within the oscillation envelope, even in the heavily crowded Galactic plane. We detect solar-like oscillations in 98 targets, including 43 with $\rm M_* > 3\ M_{\odot}$. Our sample also includes 10 stars having masses greater than $5\ \rm{M}_{\odot}$, among the highest-mass solar-like oscillators detected to date. From our detections, we find that the APOGEE DR19 spectroscopic $\log g$ is systematically larger by, on average, 0.23 dex compared to the seismic $\log g$. This offset is possibly due to the lack of intermediate-mass giants observed by \textit{Kepler}, which was used to calibrate the spectroscopic $\log g$ in the APOGEE pipeline. Extending the same pre-selection criteria to TESS targets with Gaia XP spectroscopic parameters identifies up to 37,000 candidate intermediate-mass solar-like oscillators for follow-up and population studies.
\end{abstract}


\section{Introduction} \label{sec: Intro}

Asteroseismology has become a cornerstone of stellar astrophysics, particularly for red giant branch (RGB) and red clump (RC) stars, by providing precise measurements of fundamental stellar parameters such as mass, radius, and age. This progress has been made possible by high-cadence, space-based photometric missions such as the Convection, Rotation and planetary Transits satellite (CoRoT) \citep{Baglin2006}, \textit{Kepler} \citep{Borucki2010}, K2 \citep{Howell2014}, and most recently, the Transiting Exoplanet Survey Satellite (TESS) \citep{TESS2015}. These missions have enabled the detection of solar-like oscillations in tens of thousands of RGB and RC stars \citep{Yu2018, Hon2021, Pinsonneault2025}, enriching our understanding of stellar populations across the Galaxy.

Low-mass evolved stars ($\lesssim 2\ \rm{M}_\odot$) dominate asteroseismic samples from \textit{\textit{Kepler}} \citep{Yu2018, Pinsonneault2025} and TESS \citep{Hon2021}, whereas evolved intermediate-mass stars ($\gtrsim 3\ \rm{M}_\odot$) are underrepresented despite their crucial role in constraining stellar evolution at higher-masses. Unlike their lower-mass counterparts, intermediate-mass stars ignite helium in non-degenerate cores, bypassing the helium flash, and subsequently evolve through distinct pathways on the RGB and in the core-helium-burning RC phase \citep{Girardi1999}. Their rapid evolution across these stages means that they are less frequently observed in large surveys, further contributing to their scarcity in seismic catalogs. Yet these stars serve as vital links between low-mass stars and the progenitors of massive stars, making them central to understanding the transition from long-lived, low-mass stellar populations to short-lived, high-mass populations that end as supernovae.

The internal structure of intermediate-mass stars is shaped by processes that differ substantially from those of lower-mass stars. Convective core overshoot during the main-sequence phase extends their lifetimes and alters the subsequent growth of the helium core \citep{Claret2016}, leaving imprints on later evolutionary stages \citep{Lindsay2024}. Rotational mixing is expected to be stronger in these stars, given their typically higher initial rotation rates, and angular momentum transport between the core and envelope remains poorly understood \citep{Costa2019}. In addition, the excitation and damping of solar-like oscillations in stars approaching $5\ \rm{M}_\odot$ may differ significantly from the well-studied case of low-mass giants, complicating both the detectability and interpretation of seismic signals \citep{Mosser2012, Yu2018, Srenivas2024}. For these reasons, intermediate-mass giants represent key laboratories for testing stellar models.

Asteroseismology offers a uniquely powerful method for probing these stars, as oscillation frequencies can directly reveal core properties, envelope structure, and internal rotation profiles---quantities inaccessible to traditional spectroscopic and photometric observations \citep{Beck2012, Deheuvels2012, Stello2016}. By providing precise masses, radii, and ages, seismic measurements enable robust constraints on evolutionary timescales and on the relative importance of internal mixing processes. Extending asteroseismology to intermediate-mass evolved stars would not only advance stellar physics, but also provide new insights into the demographics of evolved stellar populations in the Galaxy. However, despite these clear motivations, solar-like oscillations have proven difficult to detect in stars more massive than $\sim 5\ \rm{M}_\odot$, and no detections have yet been confirmed \citep{Crawford2024, Crawford2025}. 

This gap in the asteroseismic census stems in large part from the observational design of \textit{Kepler} \citep{Yu2018}. The APOKASC catalogs \citep{Pinsonneault2014,Pinsonneault2018,Pinsonneault2025}, formed by cross-matching \textit{Kepler} asteroseismic detections with high-resolution spectroscopy from APOGEE \citep{Majewski2017}, have yielded precise stellar parameters for nearly 16,000 RGB and RC stars. However, \textit{Kepler}'s fixed field of view, located above the Galactic plane at a galactic latitude of approximately $+13.3 \degree$, inherently limits its sampling of more massive evolved stars, which are more concentrated in the Galactic plane.

TESS, with its all-sky coverage and inclusion of the Galactic plane, presents an opportunity to extend asteroseismology to this sparsely sampled regime. Its short-cadence photometry is sensitive to oscillations in a wide range of evolved stars, including those more massive than stars previously sampled by \textit{Kepler}. However, TESS's large pixels and higher noise compared to \textit{Kepler}, make the detection of asteroseismic signals difficult in the more crowded Galactic plane. 

In this work, we demonstrate that careful light curve extraction---specifically through the manual selection of photometric apertures from TESS full-frame images---enables the detection of solar-like oscillations in evolved intermediate mass stars. We also combine spectroscopic data from APOGEE DR19 \citep{APOGEEdr192025} with asteroseismic data to infer stellar masses and radii. In Section \ref{sec: Data Preparation and Analysis}, we outline the data we use, sample selection, light curve extraction, and the estimation of seismic and stellar parameters. Section \ref{sec: Results} outlines the results of this study, presenting both a list of likely intermediate-mass giants through only spectroscopic data and a list of intermediate-mass giants with asteroseismic constraints. In Section \ref{sec: Discussion}, we discuss the systematics of the sample, low-mass contamination, and the potential for broader TESS studies of intermediate-mass giants. Lastly, we summarize our results and discuss future work in Section \ref{sec: Conclusion}.

\section{Data Preparation and Analysis} \label{sec: Data Preparation and Analysis}
Our analysis proceeds in three stages. First, we compile stellar parameters and photometric data from large-scale surveys, combining astrometry from Gaia \citep{GaiaMission2016}, high-resolution spectroscopy from APOGEE \citep{Majewski2017}, and time-series photometry from TESS \citep{TESS2015, TESS2018} (Section \ref{sec: Data Sources}). Next, we identify evolved intermediate-mass candidates from the APOGEE catalog using a combination of HR diagram cuts and spectroscopic mass estimates (Section \ref{sec: Target Selection}). For a selected subset of these stars, we generate custom TESS light curves to optimize the recovery of asteroseismic signals (Section \ref{sec: Light Curve Handling}). We then compute power spectra for these targets and measure global seismic properties (Section \ref{sec: Asteroseismic Analysis}).

\subsection{Data Sources} \label{sec: Data Sources}
Our study combines data from three primary sources:
\begin{itemize}
    \item Astrometry and broad-band photometry from Gaia --- We use Gaia DR3 \citep{GaiaDR3_2023} astrometric solutions, including parallaxes and their uncertainties, as part of our target selection process outlined in Section \ref{sec: Target Selection}. We also use distances from \citealt{Bailer-Jones2023} in conjunction with Gaia G-band photometry and extinction from the Bayestar19 dust map \citep{Green2019}, where the Bayestar19 $E(B-V)$ values are converted to Gaia Photometry following the Gaia eDR3 extinction law\footnote{\url{https://www.cosmos.esa.int/web/gaia/edr3-extinction-law}} \citep{Riello2021}, to calculate the absolute magnitude $M_G$. We then determine stellar luminosity, $L$, with the equation, 
    \begin{equation}\label{eq:luminosity}
    \frac{L}{\rm{L}_{\odot}} = 10^{-0.4(M_{\rm{G}}+BC_{\rm{G}}-\rm{M}_{\rm{bol,\odot}})}
    \end{equation}
    where $BC_{\rm{G}}$ is the bolometric correction and $\rm{M}_{\rm{bol,\odot}}$ is the Solar bolometric correction which we take to be $\rm{M}_{\rm{bol,\odot}}=4.74$ \citep{Cox2000}. We determine $BC_{\rm{G}}$ by interpolating over the MIST bolometric correction grid from the \texttt{isochrones} python package \citep{Isochrones}. 

    \item High-resolution spectroscopy from APOGEE --- Stellar parameters, including effective temperature (\teff), surface gravity ($\log g$), and metallicity ([Fe/H]) are obtained from the APOGEE DR19 catalog \citep{APOGEEdr192025}. APOGEE's infrared spectra provide precise stellar parameters for evolved stars, which we use as part of our selection process.

    \item Time-series photometry from TESS --- TESS's near all-sky coverage, 27-day sectors, and high cadence observations allow us to probe oscillations in evolved stars over a broad range of frequencies.

    \item Gaia XP spectra --- We use spectroscopic parameters from \citet{Andrae2023} in order to get an all-sky estimate for the number of potential intermediate-mass stars observed by TESS with detectable solar-like oscillations. Spectroscopic parameters are determined by the XGBoost algorithm, after training it on APOGEE spectroscopic data. In particular, we use the $\log g$ and \teff\ estimates in order to determine detection probabilities of TESS red giants that are potentially of intermediate-mass as described in Section \ref{sec: Target Selection}.

\end{itemize}

\subsection{Target Selection} \label{sec: Target Selection}
With low-mass giants being far more populous than their intermediate-mass cousins, we need to begin by isolating this sample. To do this, we utilized two tools: the HR diagram position and a spectroscopic mass estimate. We leverage the fact that with increasing mass --- at a given metallicity --- evolved stars are hotter and more luminous. We further restrict our sample by using the spectroscopic mass as an initial estimate for selection of intermediate-mass stars.

We selected target giants from the APOGEE DR19 catalog by applying cuts at
$\log g < 3.5\,\rm{dex}$ and $T_{\rm{eff}} < 6000\,\rm{K}$.
We further imposed quality cuts requiring a parallax signal-to-noise ratio
$\varpi/\sigma_{\varpi} > 10$, an effective temperature uncertainty
$\sigma_{T_{\rm{eff}}} < 200\,\rm{K}$, and a surface gravity
uncertainty $\sigma_{\log g} < 0.2$\,dex.

\begin{figure}[hbt!]
\includegraphics[width = 1.0\linewidth]{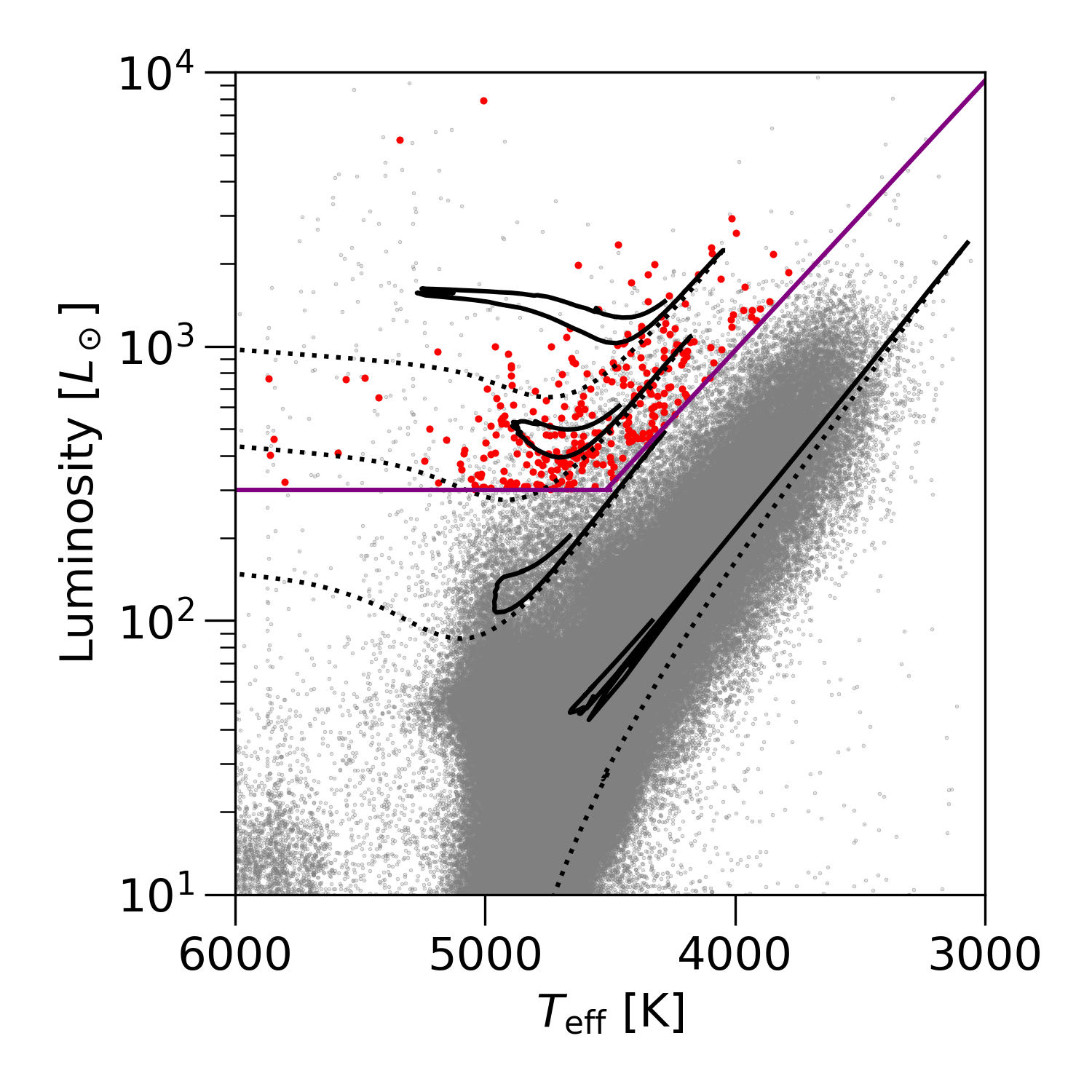}
\caption{HRD of the selection process described in Section \ref{sec: Target Selection}. The RGB selection is shown in grey and stars selected by HRD position and spectroscopic mass are shown in red. The solid purple line shows where we made our HRD selection. The evolutionary tracks represent masses of $5\rm{M}_\odot$, $4\rm{M}_\odot$, $3\rm{M}_\odot$, and $1\rm{M}_\odot$ (from top to bottom) all at solar metallicity. The solid black lines correspond to the core helium burning (CHeB) phase, the dotted black lines correspond to post main sequence, but pre-CHeB phases, and the dashed black lines correspond to the asymptotic giant branch phase.}
\label{fig:HRD_cut.png}
\end{figure}

This results in a total sample of 327,872 evolved stars to select from. We then implement a cut in Luminosity-\teff\ space near the $4\ \rm{M}_\odot$ MIST evolutionary track \citep{Choi2016} at solar metallicity (see Figure \ref{fig:HRD_cut.png}), in order to bias our sample to higher-mass stars. To remove contamination from low-mass, low-metallicity stars, we calculate spectroscopic masses from Gaia radii using the following equations:
\begin{equation}\label{eq:rad_from_lum}
    \frac{R}{\rm{R}_\odot} = \left( \frac{L}{\rm{L}_\odot} \right)^{1/2} \left( \frac{T_{\rm{eff}}}{\rm{T}_{\rm{eff,\odot}}} \right)^{-2}
\end{equation}
\begin{equation}\label{eq:mass_from_g}
    \frac{M}{\rm{M}_\odot} = \frac{g}{\mathrm{g}_\odot} \left( \frac{R}{\rm{R}_\odot} \right)^{2}
\end{equation}
We then select stars with a spectroscopic mass in the range $3.5\ \rm{M}_\odot < M_* < 10\ \rm{M}_\odot$.  This results in a sample of 300 stars, all of which are in the TESS Input Catalog \citep{Stassun2019, Paegert2021}. We then require that these stars have more than five TESS sectors to ensure sufficient temporal coverage to detect solar-like oscillations. These selections result in a sample of 227 stars\footnote{Targets were selected with a now outdated pre-release APOGEE catalog. All calculations were performed using the most up-to-date data.}. 

\begin{figure*}[hbt!]
\includegraphics[width = 1.0\linewidth]{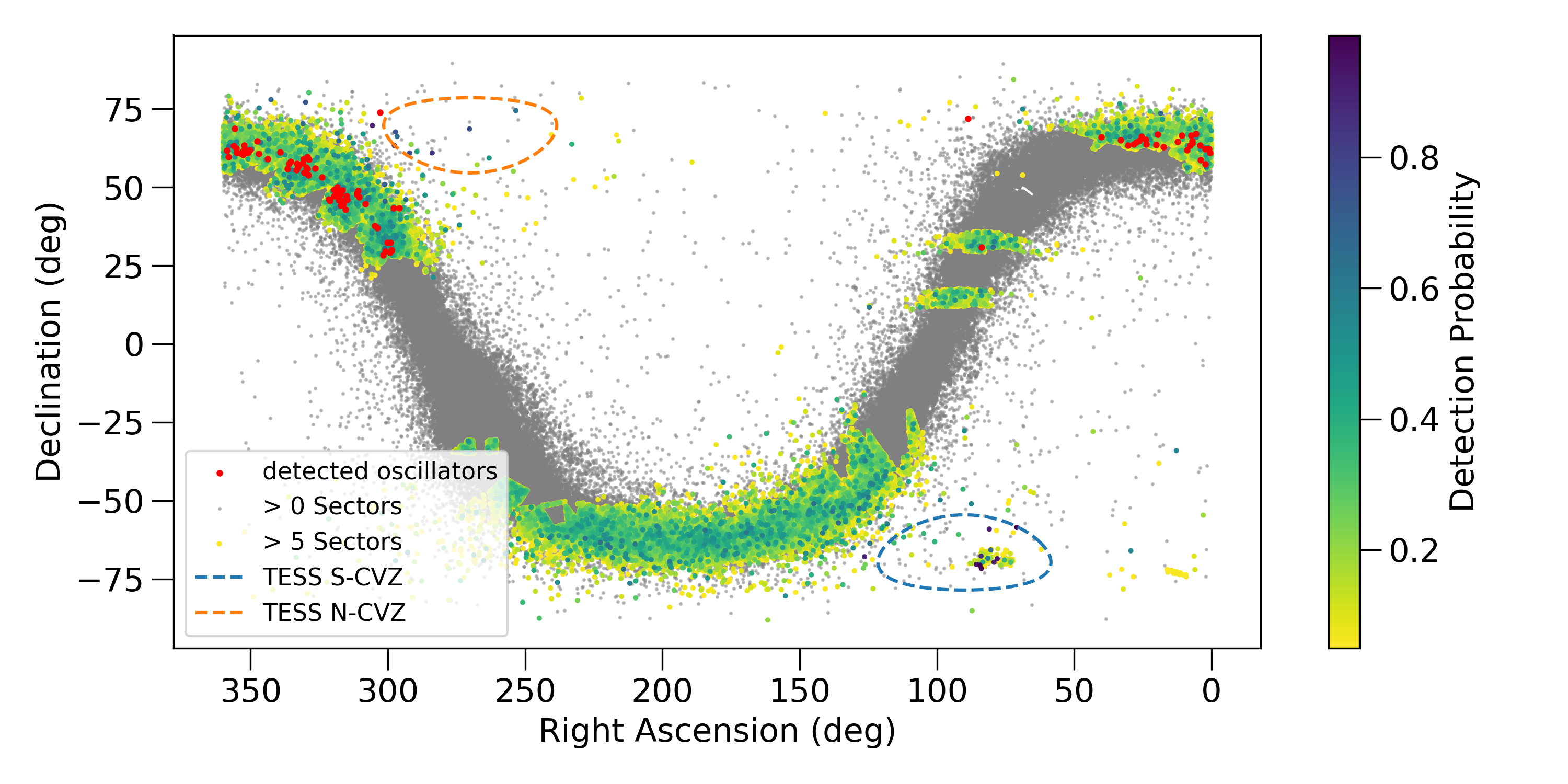}
\caption{Sky map of the TIC and \citealt{Andrae2023} catalog cross-match after applying the selection criteria outlined in Section \ref{sec: Target Selection}. Points are colored by detection probability. Grey points are the cross-match without including the criterion that targets have more than 5 sectors of TESS data. Red points are the 98 stars for which we detect oscillations as discussed in Section \ref{sec: Results}. The TESS Northern and Southern Continuous Viewing Zones are shown by orange and blue dashed lines respectively.}
\label{fig:skymap.png}
\end{figure*}

\subsection{Light Curve Handling} \label{sec: Light Curve Handling}
Adapting the methods of \citet{Saunders2022} and \citet{Stello2022a}, we generate light curves from TESS Full Frame Images downloaded from the Mikulski Archive for Space Telescopes (MAST). The 11x11 pixel cutouts are retrieved using \texttt{TESScut} \citep{Brasseur2019} from the \texttt{lightkurve} python package \citep{Lightkurve2018}. From the target pixel cutouts, we define an aperture mask by selecting the pixels with flux from the target, avoiding light from nearby stars through visual inspection. Because our targets lie in the crowded Galactic plane (see Figure \ref{fig:skymap.png}), it is difficult to ensure complete removal of contaminating flux in all cases, and such stars are flagged as potentially contaminated. 

We create an uncorrected light curve by summing the flux contained within the aperture mask at each cadence. To remove the scattered light from Earth and Moon shine, along with other instrumental systematics, we construct a design matrix from the pixels outside the aperture mask. These pixels are expected to contain little flux from the target star and therefore primarily trace background variations and detector systematics. Each column of the design matrix corresponds to the flux time series of an individual non-aperture pixel, allowing the matrix to capture background signals across the cutout. Since many of these pixel time series are highly correlated, we perform a Principal Component Analysis (PCA) on the design matrix using seven principal components. These components provide a compact basis that captures the dominant systematic variability while reducing dimensionality. 

The light curve and the reduced design matrix are then passed through \texttt{RegressionCorrector} function from \texttt{lightkurve}, which models and removes trends correlated with the principal component vectors. The resulting corrected light curve is mostly cleaned of scattered-light contamination and other instrumental effects. We found that not all noise sources were removed through this process, so each light curve was inspected by eye and any time segments with unusually large scatter were removed.

We stitch the processed sector data into one light curve for each target. We adjust observational time stamps to remove large gaps between sectors, since the presence of gaps can lead to distortions in the power spectra used for asteroseismic analysis (e.g. \citealt{Hekker2010b, Nielsen2022, GonzalezCuesta2023, Beck2026}). We calculate the Lomb-Scargle power spectral density periodogram \citep{Lomb1976, Scargle1982} which we use for our asteroseismic analysis in Section \ref{sec: Asteroseismic Analysis}.

To test the fidelity of the above methods, we also generated light curves with the Massachusetts Institute of Technology Quick-Look Pipeline (\citealt{QLPa, QLPb}, QLP) and further processed them with the PyTADACS-S (Python TESS Automated Data Analysis and Correction Software for Seismology, \citealt{Garcia2024}; Palakkatharappil et al. in prep.) based on the KADACS (Kepler Automated Data Analysis and Correction Software) developed for the \emph{Kepler} mission \citep{Garcia2011, Garcia2014, Pires2015}. First, the pipeline stitches together consecutive sectors when the gap between them is no more than three sectors. Then it applies two stages of sigma clipping. In the first stage, any data points with flux values more than $10\sigma$ from the mean flux are removed. In the second stage, points are removed if they deviate more than $4\sigma$ from the original light curve after being smoothed with a median filter of 1-day width. The light curve is then binned into a 30-minute cadence via the nearest-neighbor resampling algorithm with the slotting principle, as described in \citet{Garcia2014}. The final step is a high-pass filter using a triangular kernel with a 55-day width.

To compare the QLP data to our custom light curves, we measure a power-to-background ratio (PBR) within the oscillation envelope of the power spectrum. To do this, we first compute the power spectrum of the light curves for all targets where we detect solar-like oscillations. We smooth the power spectrum with a gaussian filter of width equal to \dnu, where \dnu\ is given by the \numax-\dnu\ relation from \citet{Stello2009}. To determine a global background---covering both granulation at low frequencies and white noise at high frequencies---we remove the power excess within $\pm3\ \Delta\nu$ around \numax\ and interpolate over the gap with a linear background model in log space. We then calculate the PBR with the following equation:
\begin{equation}\label{eq:PBR}
    \rm{PBR} = \frac{1}{N}\sum_{i=1}^N \frac{P_i}{B_i}
\end{equation}
where $N$ is the number of frequency bins in the power spectrum, $P$ is the smoothed power spectrum, and $B$ is the global background model.

\subsection{Asteroseismic Analysis} \label{sec: Asteroseismic Analysis}

Asteroseismology can infer stellar parameters by using two measurable quantities: \numax, the frequency of maximum power, and \dnu, the average frequency spacing between modes of the same spherical harmonic degree $\ell$. The seismic parameter \numax\ is related to $\log g$ and \teff\ \citep{Brown1991, Kjeldsen1995, Belkacem2011, Hekker2020} and \dnu\ is related to the mean density \citep{Ulrich1986} through the following relations:
\begin{equation}\label{eq:numax_scaling}
    \frac{g}{\mathrm{g}_\odot} = \frac{\nu_{\rm{max}}}{\nu_{\rm{max,\odot}}} \left( \frac{T_{\rm eff}}{\rm{T}_{\rm{eff},\odot}} \right)^\frac{1}{2}
\end{equation}
\begin{equation}\label{eq:dnu_scaling}
\frac{\Delta\nu}{\Delta\nu_\odot} = \sqrt{\frac{\Bar{\rho}}{\Bar{\rho}_\odot}}
\end{equation}
These relationships can be combined into the following seismic scaling relations to infer stellar radius and mass (hereafter referred to as the double-scaling relation, or DSR):
\begin{equation}\label{eq:rad_dsr}
   \frac{R}{\rm{R}_\odot} = \left(\frac{\nu_{\rm max}}{f_{\nu_{\rm max}}\nu_{\rm max,\odot}}\right) \left(\frac{\Delta\nu}{f_{\Delta\nu}\Delta\nu_{\odot}} \right)^{-2} \left( \frac{T_{\rm eff}}{\rm{T}_{\rm{eff},\odot}} \right)^{\frac{1}{2}}
\end{equation}
\begin{equation}\label{eq:mass_dsr}
   \frac{M}{\rm{M}_\odot} = \left(\frac{\nu_{\rm max}}{f_{\nu_{\rm max}}\nu_{\rm max,\odot}}\right)^3 \left(\frac{\Delta\nu}{f_{\Delta\nu}\Delta\nu_{\odot}} \right)^{-4} \left( \frac{T_{\rm eff}}{\rm{T}_{\rm{eff},\odot}} \right)^{\frac{3}{2}}
\end{equation}
With an independent radius measurement one can instead infer mass using the following equation (hereafter referred to as the single-scaling relation, or SSR):
\begin{equation}\label{eq:mass_ssr}
    \frac{M}{\rm{M}_\odot} = \left(\frac{\nu_{\rm max}}{f_{\nu_{\rm max}}\nu_{\rm max,\odot}}\right) \left( \frac{T_{\rm eff}}{\rm{T}_{\rm{eff},\odot}} \right)^{1/2} \left( \frac{R}{\rm{R}_\odot} \right)^2
\end{equation}
where $f_{\nu_{\rm max}}$ and $f_{\Delta\nu}$ are correction factors designed to account for offsets between fundamental and asteroseismic parameters. $f_{\nu_{\rm max}}$ is an empirical calibration anchored to some alternative measurement of mass or radius. Some examples included using Gaia radii \citep{Pinsonneault2025}, or seismic masses and radii from individual frequency modeling \citep{Huber2024, Lindsay2026}, as the basis. $f_{\Delta\nu}$ is computed theoretically from stellar models \citep{White2011, Sharma2016asfgrid, Stello2022b}.

To measure \numax, we inspect the power spectra computed from the light curves discussed in Section \ref{sec: Light Curve Handling} by eye, using the spectroscopic $\log g$ as a guide. We then take our by-eye \numax\ estimate and pass it into \texttt{pyMON}\footnote{\url{https://github.com/maddyhowell/pyMON}} \citep{Howell2025}, which adapts the \numax\ detection methods of \texttt{pySYD} \citep{Huber2009, Chontos2022}. We choose \texttt{pyMON} since it is well-suited for detecing signals in giants with both low-\numax\ and low-signal-to-noise---as is expected to be the case in our intermediate-mass sample \citep{Yu2018, Crawford2024}. Using the initial \numax, \texttt{pyMON} defines a power excess window. For particularly noisy targets, we manually defined the upper and lower boundaries of the power excess window. Then a linear background model is fit in log space to the region of the power spectrum within the defined power-excess window and subtracted from the spectrum. By fitting a linear background model, \texttt{pyMON} is able to bypass a failure mode of \texttt{pySYD}, which arises from its inability to fit a Harvey-like background to low-\numax\ giants. The spectrum is then heavily smoothed using the \dnu\ estimate provided by the \numax–\dnu\ scaling relation from \citealt{Stello2009}. After smoothing, the frequency of maximum amplitude is adopted as our \numax\ measurement, with the uncertainty estimated from the standard deviation of 500 realizations of this procedure, each generated with stochastic noise. Since intermediate-mass stars spend the majority of their post-main-sequence life in the core-helium-burning phase, we assume their $f_{\nu_{\rm{max}}}$ correction factor to be of order unity, consistent with previous studies \citep{Yu2018, Zinn2019, Li2023, Crawford2024}. Our determination of \dnu\ and $f_{\Delta\nu}$ is outlined in Appendix \ref{sec:appendix}, along with a discussion of the DSR seismic masses and example power spectra (see Figure \ref{fig:example_psd.png}). 

\begin{deluxetable*}{lccccccccc}
\tablecaption{Sample of the asteroseismic detections.}
\label{tab:catalog_sample}
\tablehead{
\colhead{TIC ID} &
\colhead{$\rm{N_{Sectors}}$} &
\colhead{$\rm{T_{mag}}$} &
\colhead{$\nu_{\max}$} &
\colhead{$R_{\rm{Gaia}}$} &
\colhead{$T_{\rm{eff}}$} &
\colhead{$M_{\rm{spec}}$} &
\colhead{$M_{\rm{SSR}}$} &
\colhead{Cont.} \\
&
&
&
\colhead{($\mu$Hz)} &
\colhead{($\rm{R}_\odot$)} &
\colhead{($\rm{K}$)} &
\colhead{($\rm{M}_\odot$)} &
\colhead{($\rm{M}_\odot$)} &
&
}
\startdata
365033429 & $9$ & $11.793$  & $14.71 \pm 0.24$ & $41.28 \pm 2.14$ & $4402.75 \pm 50$ & $9.84 \pm 1.03$ & $7.08 \pm 0.12$ & 0 \\
10242756  & $8$ & $12.517$  & $48.41 \pm 0.61$ & $20.79 \pm 1.63$ & $5474.74 \pm 50$ & $5.19 \pm 0.81$ & $6.59 \pm 0.09$ & 0 \\
12703731  & $7$ & $11.441$  & $37.92 \pm 0.62$ & $23.05 \pm 1.78$ & $5105.03 \pm 50$ & $7.79 \pm 1.21$ & $6.13 \pm 0.10$ & 1 \\
195727782 & $9$ & $11.57$   & $8.94 \pm 0.22$  & $48.15 \pm 2.08$ & $4350.58 \pm 50$ & $5.68 \pm 0.52$ & $5.82 \pm 0.14$ & 1 \\
374030825 & $7$ & $10.644$  & $4.76 \pm 0.13$  & $63.21 \pm 4.60$ & $4474.69 \pm 50$ & $9.42 \pm 1.40$ & $5.41 \pm 0.16$ & 0 \\
\enddata
\tablecomments{
Only a subset of the full sample studied in this work is shown here for reference. The full table is available in machine-readable format online. We also only show some of our derived parameters used in this study. The full online table will contain a more exhaustive list of stellar parameters. Quoted uncertainties correspond to 1$\sigma$ errors.}
\end{deluxetable*}

\section{Results} \label{sec: Results}
\subsection{Seismic Detections and Masses}
\begin{figure}[hbt!]
\includegraphics[width=1.0\linewidth]{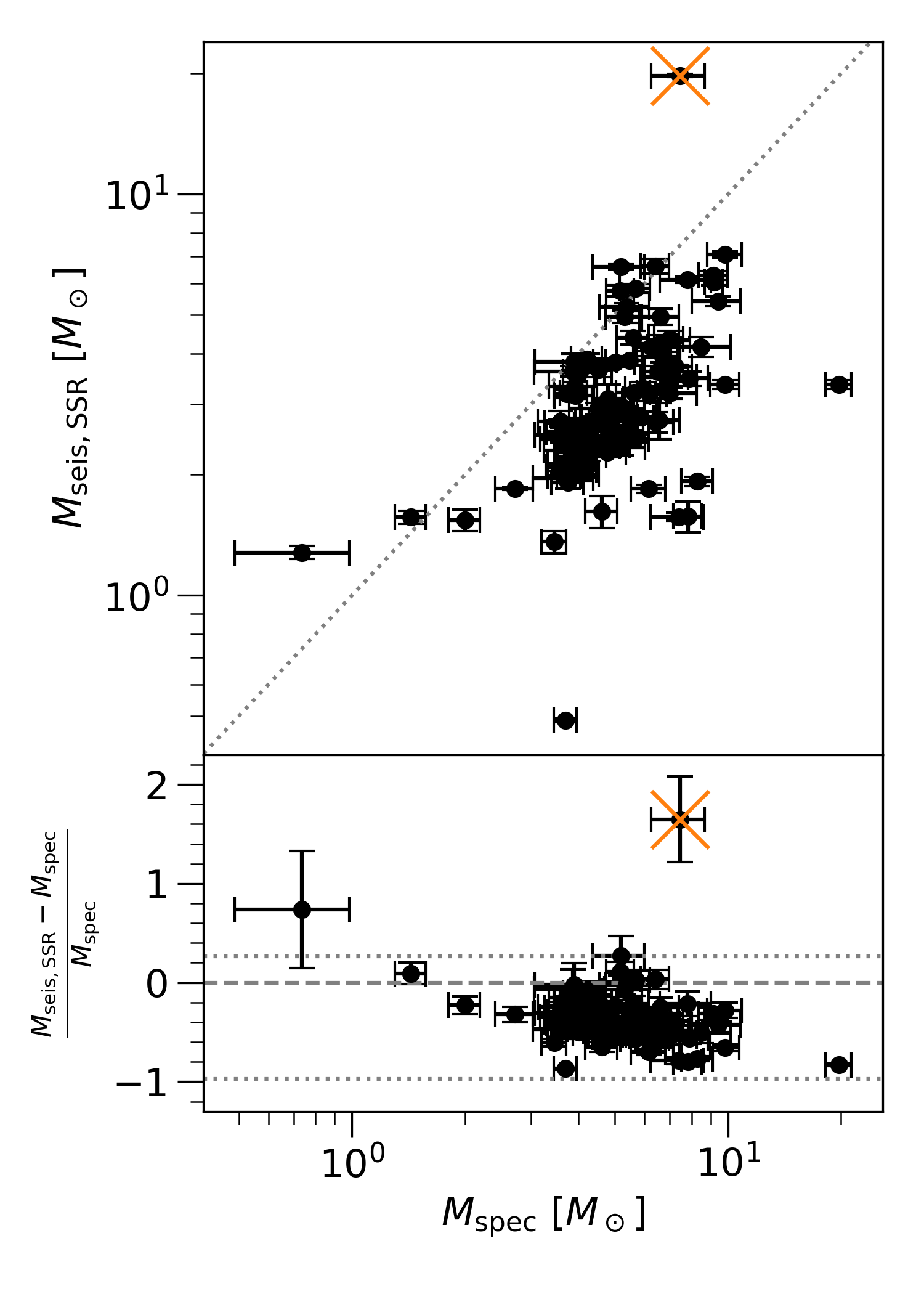}
\caption{\textbf{Top Panel:} Comparison of the SSR mass to the spectroscopic mass for all stars with measured \numax. Dotted line represents a 1:1 mass relation. \textbf{Bottom Panel:} Fractional mass residuals for all stars with measured \numax. The dashed line represents the 1:1 mass relation and the two dotted lines correspond to $\pm2\sigma$ around the mean of the residual distribution. The mean fractional residual is $\sim 0.35$. In both panels the orange marker indicates TIC 10431423, an outlier discussed in Section \ref{sec: systematics}.}
\label{fig:masses.png}
\end{figure}

From the 227 stars for which we made light curves, we found 98 exhibiting solar-like oscillations with a measurable \numax. Of the 98 stars with detected oscillations, we flagged 45 of them as potentially contaminated by analyzing TESS full-frame images as discussed in Section \ref{sec: Light Curve Handling}. We further measured \dnu\ for 71 stars, 31 of which are flagged as potentially contaminated, and discuss these measurements in Appendix \ref{sec:appendix}. We show a comparison between the spectroscopic masses and the SSR seismic masses in Figure \ref{fig:masses.png}. There is a notable difference between both mass measurements, with seismic masses on average $~35\%$ lower than spectroscopic masses. We attribute this to an offset in seismic and spectroscopic $\log g$ shown in Figure \ref{fig:logg_offset.png} and discussed in Section \ref{sec: systematics}. Using the SSR mass estimates, we find a total of 44 intermediate-mass stars ($3\ \rm{M}_{\odot} < M_* < 8\ \rm{M}_{\odot}$). Of those intermediate-mass stars, we find 10 with masses greater than $5\ \rm{M}_\odot$, \textit{making these some of the highest-mass stars with observed solar-like oscillations}. We list a few of the highest-mass stars we find in Table \ref{tab:catalog_sample} for reference.

\subsection{Light Curve Validation}\label{sec: Light Curve Validation}

\begin{figure}[hbt!]
\includegraphics[width=1\linewidth]{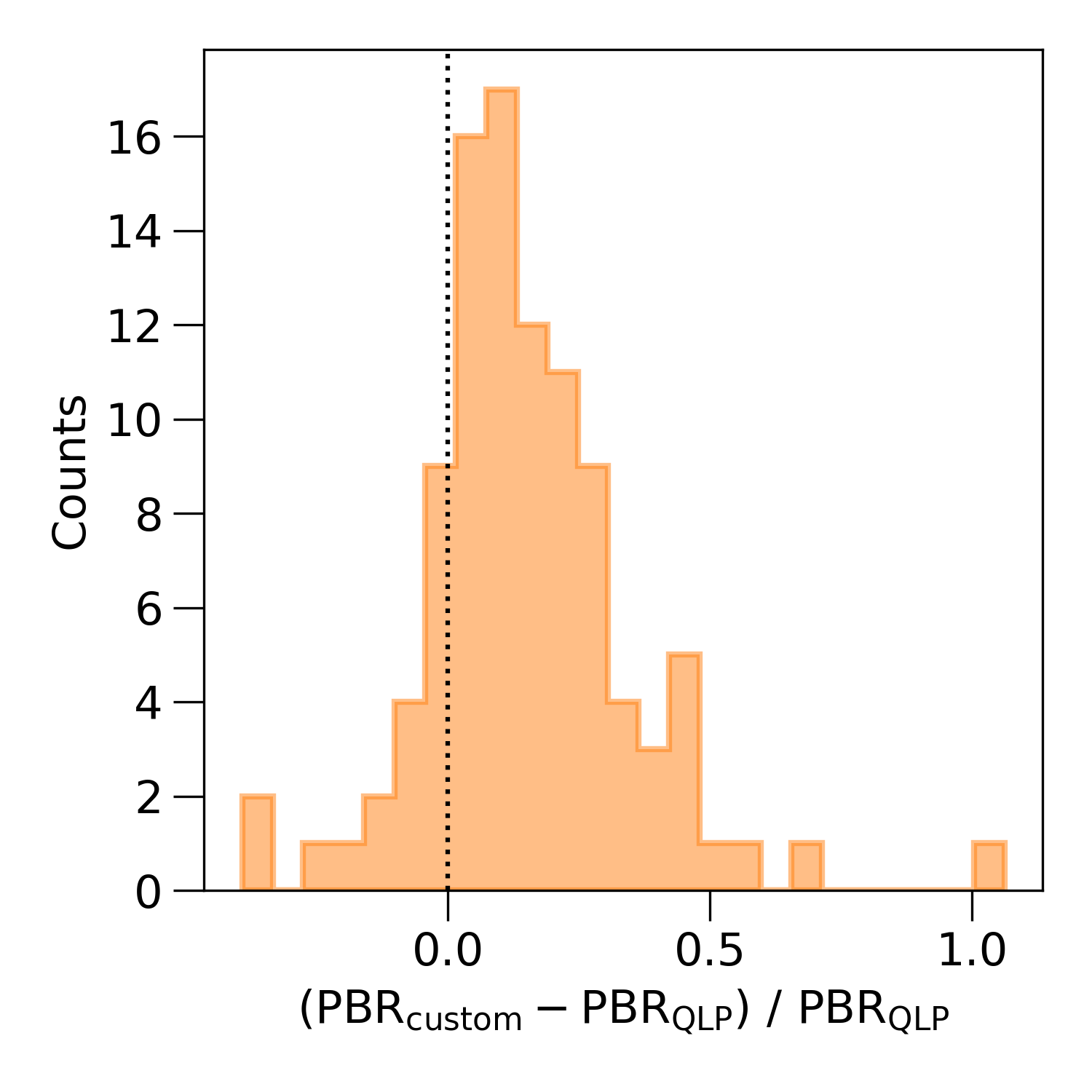}
\caption{The fractional difference between the PBR of the custom power spectra of our detected oscillators compared to the PBR of QLP power spectra for the same targets. Positive values indicate better performance with our custom power spectra. The dotted black line marks where the PBR is the same in both power spectra. The median of the distribution is a 12\% increase in $\rm{PBR_{custom}}$ relative to $\rm{PBR_{QLP}}$.}
\label{fig:snr_hist.png}
\end{figure}

\begin{figure}
\includegraphics[width=1\linewidth]{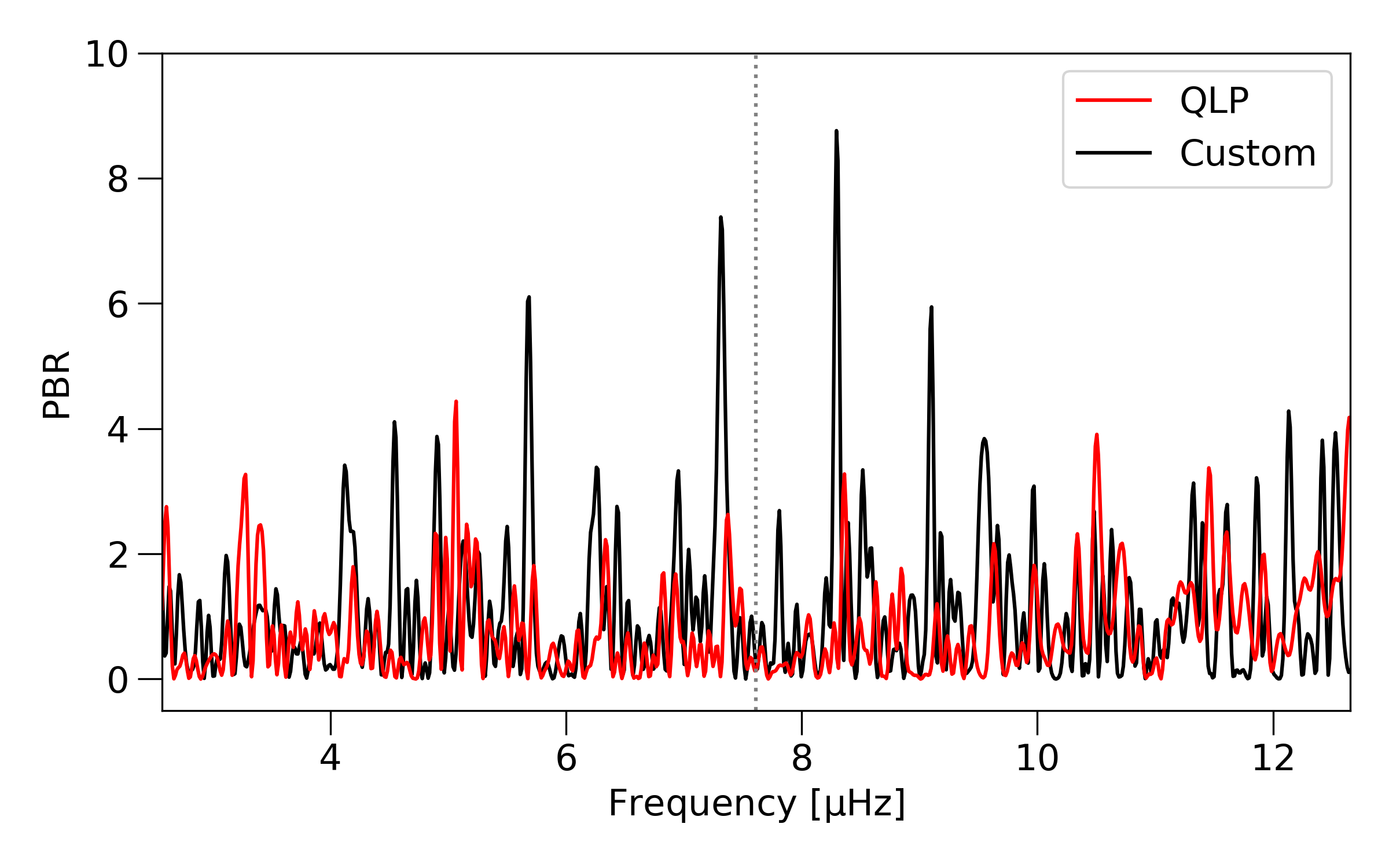}
\caption{Background-divided power spectra for TIC 11114629 from QLP (red) and our custom aperture photometry (black), shown without smoothing. The measured \numax\ is denoted by the dotted black line. The frequency range shown is $\pm4\ \Delta\nu$ around \numax\, where \dnu\ is estimated using the \numax-\dnu\ relation from \citet{Stello2009}.}
\label{fig:bg_div_ps.png}
\end{figure}

For most targets, we observe an increase in the PBR of the oscillation power excess when using our custom light curves relative to QLP, as shown in Figure \ref{fig:snr_hist.png}. The distribution is skewed toward positive values, with a median increase of 12\% in the PBR for the custom reductions relative to QLP. This indicates that the oscillation signal is recovered with greater contrast against the stellar and instrumental background. Figure \ref{fig:bg_div_ps.png} illustrates this effect for TIC 11114629. In the background-divided power spectrum, the oscillation excess near \numax\ is substantially more prominent in the custom reduction. In contrast, the QLP spectrum exhibits elevated background power, which reduces the visibility of the oscillation envelope and therefore its detectability. Overall, these results demonstrate that the custom pipeline generally enhances oscillation detectability by suppressing residual systematics and excess background noise, while only a small subset of targets show little or no improvement.

\section{Discussion} \label{sec: Discussion}

\subsection{$\log g$ Systematics and Outliers} \label{sec: systematics}

\begin{figure}[hbt!]
\includegraphics[width = 1.0\linewidth]{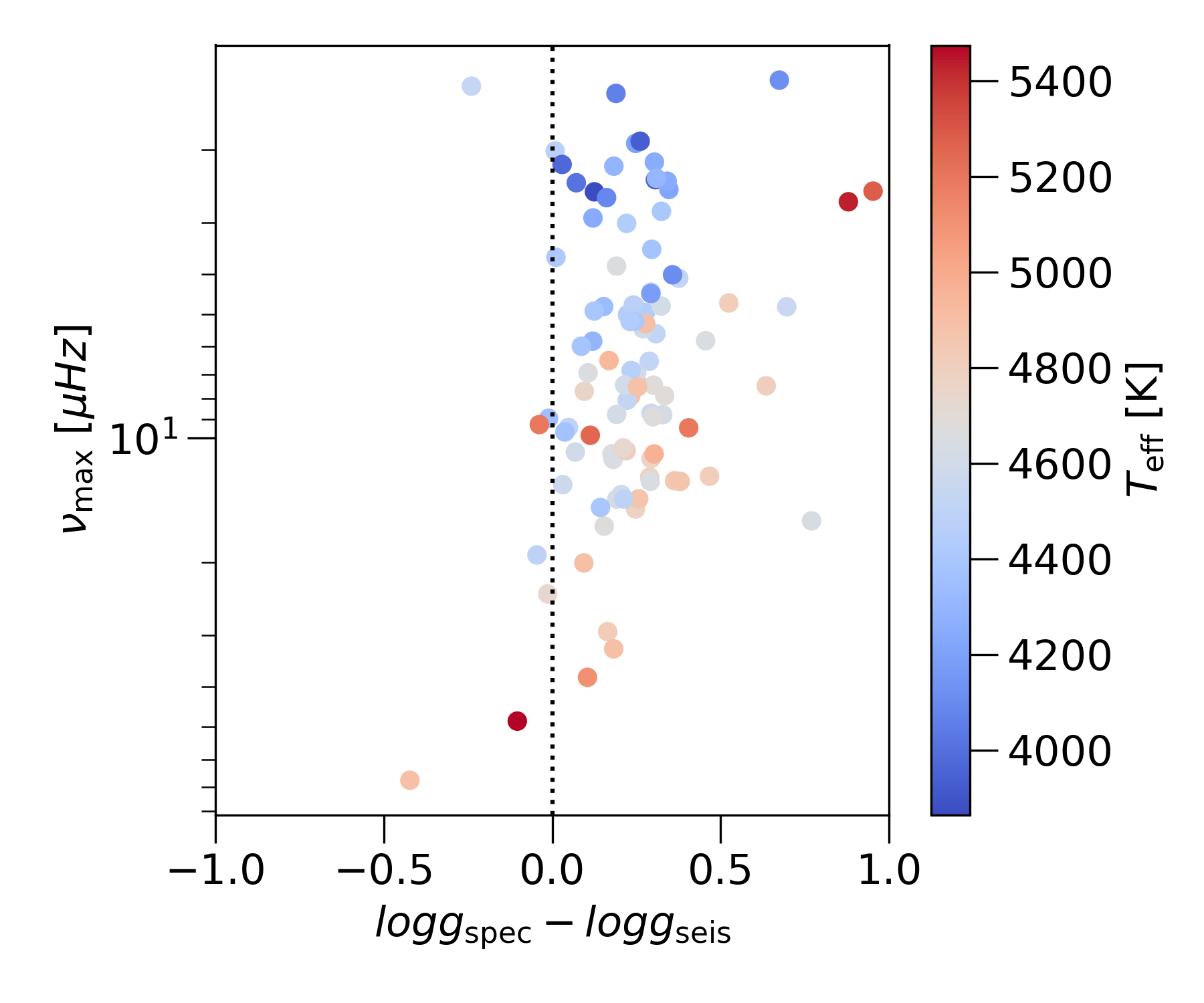}
\caption{Offset between spectroscopic and seismic $\log g$ compared to \numax. Points are colored by \teff. The dotted line corresponds to no offset.}
\label{fig:logg_offset.png}
\end{figure}

We find an average offset of $0.23\pm0.28\ \rm{dex}$ between our spectroscopic $\log g$ values and seismic $\log g$ values, as shown in Figure \ref{fig:logg_offset.png}. This discrepancy appears to vary as a function of \numax, with the offset increasing with decreasing \numax. There is also a minor trend with an increasing offset with decreasing \teff, although this effect is much smaller. This $\log g$ offset is likely caused by a misclassification in the APOGEE pipeline that leads to an incorrect $\log g$ calibration. Although these stars are in the core-helium burning phase, their high masses lead to significantly lower surface gravities compared to their lower-mass counterparts, while remaining hotter than the RGB. Since APOGEE $\log g$ values are calibrated to seismic $\log g$ values, the lack of seismic sampling in this mass regime would lead to higher $\log g$ values given the temperature range of our sample. This follows with the trends noted in \numax\ and \teff, since stars with a lower \numax\ (i.e. lower $\log g$) and a hotter \teff, would be further from the space where seismic calibrators are present leading to larger offsets. The effect of this offset manifests itself clearly in the form of the SSR seismic masses. As seen in the left panel of Figure \ref{fig:masses.png}, seismic masses are consistently lower than spectroscopic masses, since the only effective change between equation \ref{eq:mass_from_g} and equation \ref{eq:mass_ssr} is the use of a seismic $\log g$ as \numax\ relates to $\log g$ through equation \ref{eq:numax_scaling}.

There are additional sources of uncertainty in spectroscopic $\log g$ measurements that stem from limitations in stellar atmosphere modeling for red giants \citep{Asplund2005}. These systematic uncertainties can be calibrated against seismic measurements of $\log g$, which are largely model-independent \citep{Holtzman2018}. However, due to the sparse sample of existing intermediate-mass seismic calibrators, spectroscopic $\log g$ values for stars in our mass range may reflect both intrinsic modeling uncertainties and the extrapolation of empirical calibrations beyond the regime in which seismic benchmarks are available. As such, we opt to trust our measurement of \numax---and thus our seismic $\log g$---because it places tighter constraints on stellar mass. 

TIC 10431423 is a notable outlier present in Figure \ref{fig:masses.png}, due to its mass residual far beyond the $\pm2\sigma$ region. This star was flagged as contaminated, which likely explains its anomalously large seismic mass of $19.77\pm0.19,\rm{M}_\odot$. The contaminating source is likely a lower mass RGB or RC star that would have been excluded by our selection criteria. This is further supported by its large measured \numax\ ($67.31\pm0.53\ \rm{\mu Hz}$), which does not match the lower \numax\ values seen across sample. An artificially high \numax\ due to contamination would bias the mass estimate to higher-masses following equation \ref{eq:mass_ssr}.

\subsection{Low-Mass Contamination} \label{sec: Low-Mass Contamination}

Although we tried to minimize low-mass stars in our sample via the selection process described in Section \ref{sec: Target Selection}, we still find 53 low-mass giants are present when using the SSR mass estimate. Looking at Figure \ref{fig:HRD_cut.png}, it can be seen that lower-mass evolutionary tracks do not begin to contaminate our sample until they reach the asymptotic giant branch phase. The low-mass interlopers were found to have lower \numax\ values compared to the intermediate-mass sample, which would be consistent with the asymptotic giant branch or, for more metal-poor cases, the upper RGB. It is also possible that the oscillations found in these stars are due to contamination from a nearby star, as 20 of these stars were flagged as potentially contaminated. Should these stars be contaminated by more metal-poor AGB or upper RGB stars, then the false positive \numax\ detection would lead to lower surface gravities and lower mass estimates. 

\subsection{Spectroscopic Checks} \label{sec: Spectroscopic Checks}

\begin{figure}[hbt!]
\includegraphics[width = 1.0\linewidth]{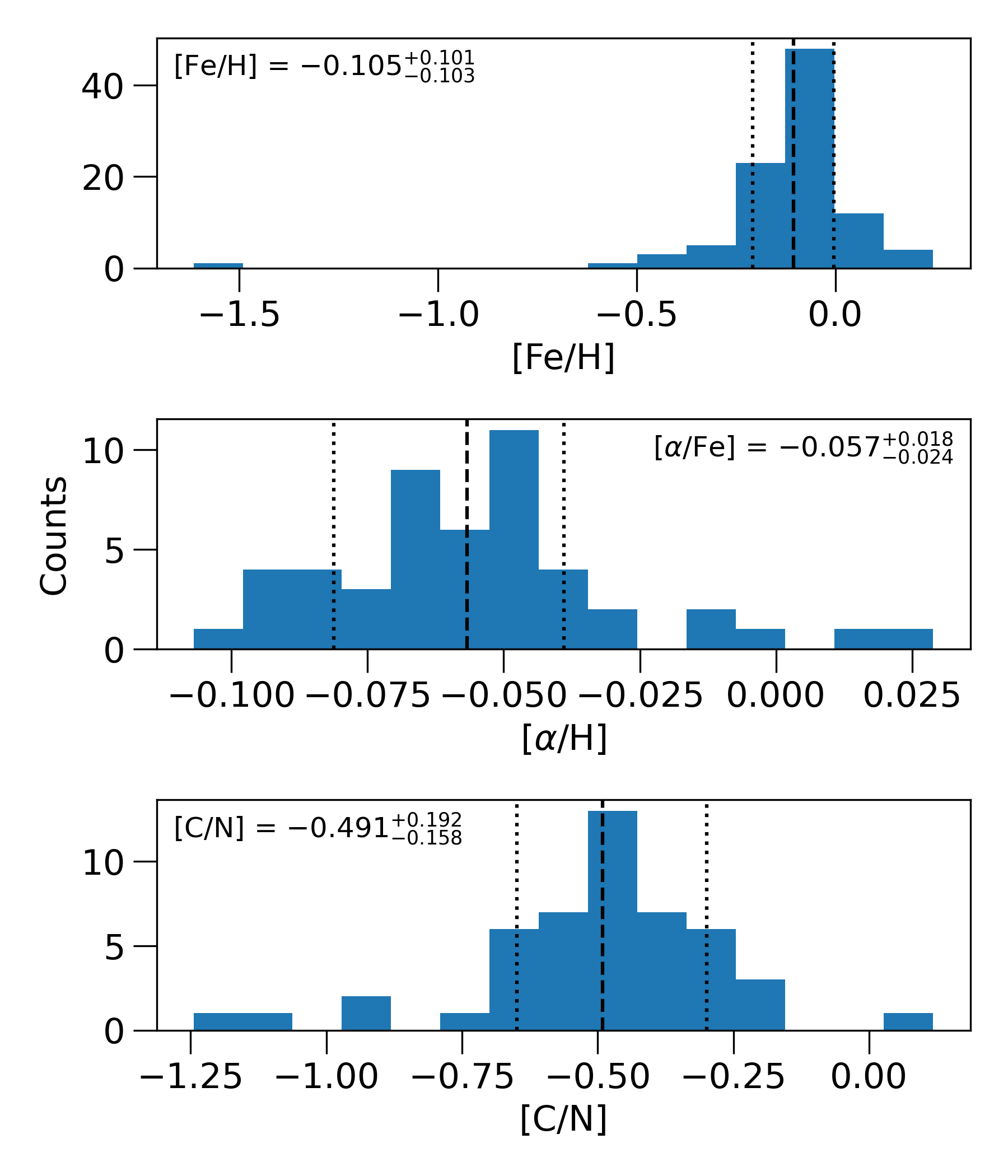}
\caption{Abundance distributions for [Fe/H], [$\alpha$/Fe], and [C/N] (top, middle, and bottom respectively). Black dotted lines correspond to the 16th and 84th percentile and the black dashed line corresponds to the 50th percentile.}
\label{fig:chemistry.png}
\end{figure}

As an additional check on the high-mass nature of our targets, we examined elemental abundance ratios from APOGEE spectra that are informative for stellar mass/age in red giants: [Fe/H], $\rm[\alpha/Fe]$, and [C/N]. For the full set of \numax\ detections, we find a median [Fe/H] of -0.105 dex, broadly consistent with a young, thin-disk population. Not all stars have reliable measurements of every abundance ratio; the discussion of $\rm[\alpha/Fe]$ and [C/N] therefore uses the subsets with available APOGEE solutions (49 stars for $\rm[\alpha/Fe]$ and 48 stars for [C/N]). The sample is predominantly sub-solar in $\rm[\alpha/Fe]$ with a median of -0.057 dex, consistent with the young, low-$\alpha$ Milky Way population \citep{Fuhrmann1998, Bensby2003}. We also find systematically low [C/N] values with a median of -0.481 dex, which has been shown to correlate with higher mass and younger ages in evolved stars \citep{Roberts2024, Roberts2026}. Distributions of all three abundance ratios are shown in Figure \ref{fig:chemistry.png}. We emphasize that these abundance ratios provide population-level support rather than a definitive mass diagnostic for individual stars, given their dependence on evolutionary state and internal mixing. 

\subsection{Potential for Future Work} \label{sec: Future Work}

We also performed the cuts from Section \ref{sec: Target Selection} on a sample of stars created by cross-matching the TESS Input Catalog \citep{Stassun2018} and the \citet{Andrae2023} catalog, allowing for full-sky coverage. This leads to a total sample of 92,183 targets. If the number of sectors is removed as a selection criterion, then the sample increases to 242,803 targets. We calculate detection probabilities for these targets using the \texttt{TESS-atl} python package \citep{Hey2024_ATLcode}, which determines detection probabilities by using apparent magnitude, \teff, radius, $\log g$, the number of observed TESS sectors, and the observing cadence as inputs. These detection probabilities, as well as the positions on the sky for these targets, are shown in Figure \ref{fig:skymap.png}. To infer the number of expected detections, we take the number of targets above a certain \texttt{TESS-atl} detection probability and multiply it by 43\%, which is the recovery rate of \numax\ for the 227 stars we study in this paper (see Section \ref{sec: Asteroseismic Analysis} and \ref{sec: Results}). We summarize these counts in Table \ref{tab: predicted yield}.

\begin{table}[hbt!]
    \centering
    \begin{tabular}{|c|c|}
        \hline
        $\rm{P}_\texttt{TESS-atl}$ & Counts \\
        \hline
        $>$ 0.05 & 36,620 \\
        $>$ 0.5 & 98 \\
        $>$ 0.6 & 22 \\
        $>$ 0.7 & 7 \\
        $>$ 0.8 & 5 \\
        $>$ 0.9 & 3 \\
        $>$ 0.95 & 1 \\
        \hline
    \end{tabular}
    \caption{Estimated number of seismic detections of TIC targets with Gaia XP spectroscopic parameters that pass the selection criteria outlined in Section \ref{sec: Target Selection} given a threshold \texttt{TESS-atl} detection probability.}
    \label{tab: predicted yield}
\end{table}

These targets with high detection probabilities provide a useful sample for finding likely intermediate-mass oscillating red giants. In particular, we point to the handful of stars found in the Southern Continuous Viewing Zone (see Figure \ref{fig:skymap.png}), which could provide enough time-series data for a more detailed seismic analysis. We also include detection probabilities above the \texttt{TESS-atl} limit of 0.05, since the \texttt{TESS-atl} probabilities are statistical estimates rather than strict detection thresholds, and oscillation signals may still be detected at lower probabilities.

\section{Conclusion} \label{sec: Conclusion}

In this paper, we constructed a candidate list of 227 evolved intermediate-mass stars and generated light curves for them by using custom photometric apertures from TESS full frame images. From their power spectra, we measured \numax\ for 98 stars, allowing us to estimate seismic masses using the single-scaling relation (equation \ref{eq:mass_ssr}) and compare them to spectroscopic masses. Taking the SSR, which combines \numax\ with a radius from Gaia, as our most precise estimator, we identified 43 intermediate-mass stars ($3\ \rm{M}_\odot \lesssim M_* \lesssim 8\ \rm{M}_\odot$), including 10 with masses exceeding $5\ \rm{M}_\odot$. 

These detections mark an important step toward extending asteroseismology to a mass regime that has been largely inaccessible to previous surveys. Measuring seismic masses for evolved intermediate-mass stars provides a new avenue for testing models of stellar structure and evolution in a critical transitional range---where convective core overshoot, rotational mixing, and internal angular momentum transport begin to shape post-main-sequence evolution in ways that differ from lower-mass red giants. The ability to detect solar-like oscillations in stars approaching and exceeding $5\ \rm{M}_\odot$ therefore offers an opportunity to empirically constrain how these processes scale with mass and influence the subsequent core-helium-burning phase. More broadly, these measurements fill a key gap between low-mass red giants that dominate current seismic catalogs and the high-mass progenitors of core-collapse supernovae, connecting two regimes of stellar evolution that have so far been studied largely in isolation.

This work only touches the surface of the future potential to study intermediate-mass giants with asteroseismology. Beyond our identification of 43 intermediate-mass red giants, we identify a target list of 92,183 stars observed by TESS with available Gaia XP spectroscopic parameters that have potential to be intermediate mass by the same selection criteria we employed for the 43 intermediate-mass stars found in this study. Looking forward, the Nancy Grace Roman Space Telescope \citep{RomanMission2025} will launch as early as August 2026 and perform a high cadence time-domain survey of the Galactic bulge. This has the potential to yield approximately 300,000 \numax\ detections in oscillating red giants, with about 1-2\% of those estimated to be intermediate-mass \citep{WeissDowning2025}.

In future work, it will be important to perform a more detailed analysis of these targets and others across the TESS fields. As noted in \citet{Pinsonneault2025}, employing multiple asteroseismic pipelines can yield more precise and consistent stellar parameters, a natural next step to validate and refine our current mass estimates. A more detailed frequency analysis, alongside improved noise modeling, could further confirm the intermediate-mass nature of these stars and probe their internal structure more deeply. Expanding this sample with additional spectroscopic surveys or ground-based follow-up would provide a broader basis for comparison with stellar evolution models. Finally, because our manual aperture selection proved highly effective in improving signal recovery, developing an automated version of this method would enable larger searches for oscillations in the most massive TESS giants, advancing the reach of population asteroseismology toward more massive stars.

\section*{Acknowledgments}
NJD \& MHP acknowledge support from the NASA grant 80NSSC24K0091. 
MH \& MHP acknowledge support from the NASA grant 80NSSC24K0637. RAG, DBP, and LB acknowledge financial support from the Centre national d’études spatiales (CNES), France (ROR: https://ror.org/04h1h0y33), within the framework of the PLATO space mission. S.M. acknowledges support from the Spanish Ministry of Science and Innovation (MICINN) with the grant No. PID2023-149439NB-C41. The authors appreciate the helpful comments and discussion from those within the APOTESS collaboration. The authors thank Jamie Tayar and Timothy Bedding for feedback on the manuscript. NJD thanks Christopher Lindsay and Sarbani Basu for their useful discussion. 

This research made use of Lightkurve, a Python package for Kepler and TESS data analysis \citep{Lightkurve2018}.

Some of the data presented in this paper were obtained from the Mikulski Archive for Space Telescopes (MAST) at the Space Telescope Science Institute. The specific observations analyzed can be accessed via \dataset[https://doi.org/10.17909/0cp4-2j79]{https://doi.org/10.17909/0cp4-2j79} and \dataset[https://doi.org/10.17909/t9-r086-e880]{https://doi.org/10.17909/t9-r086-e880}. STScI is operated by the Association of Universities for Research in Astronomy, Inc., under NASA contract NAS5–26555. Support to MAST for these data is provided by the NASA Office of Space Science via grant NAG5–7584 and by other grants and contracts.

We acknowledge the use of TESS High Level Science Products (HLSP) produced by the Quick-Look Pipeline (QLP) at the TESS Science Office at MIT, which are publicly available from the Mikulski Archive for Space Telescopes (MAST). Funding for the TESS mission is provided by NASA's Science Mission directorate.

Funding for the Sloan Digital Sky Survey V has been provided by the Alfred P. Sloan Foundation, the Heising-Simons Foundation, the National Science Foundation, and the Participating Institutions. SDSS acknowledges support and resources from the Center for High-Performance Computing at the University of Utah. SDSS telescopes are located at Apache Point Observatory, funded by the Astrophysical Research Consortium and operated by New Mexico State University, and at Las Campanas Observatory, operated by the Carnegie Institution for Science. The SDSS website is \url{www.sdss.org}.

SDSS is managed by the Astrophysical Research Consortium for the Participating Institutions of the SDSS Collaboration, including the Carnegie Institution for Science, Chilean National Time Allocation Committee (CNTAC) ratified researchers, Caltech, the Gotham Participation Group, Harvard University, Heidelberg University, The Flatiron Institute, The Johns Hopkins University, L'Ecole polytechnique f\'{e}d\'{e}rale de Lausanne (EPFL), Leibniz-Institut f\"{u}r Astrophysik Potsdam (AIP), Max-Planck-Institut f\"{u}r Astronomie (MPIA Heidelberg), Max-Planck-Institut f\"{u}r Extraterrestrische Physik (MPE), Nanjing University, National Astronomical Observatories of China (NAOC), New Mexico State University, The Ohio State University, Pennsylvania State University, Smithsonian Astrophysical Observatory, Space Telescope Science Institute (STScI), the Stellar Astrophysics Participation Group, Universidad Nacional Aut\'{o}noma de M\'{e}xico, University of Arizona, University of Colorado Boulder, University of Illinois at Urbana-Champaign, University of Toronto, University of Utah, University of Virginia, Yale University, and Yunnan University.

\software{astropy \citep{astropy13, astropy18, astropy22}, 
pandas \citep{pandas20}, lightkurve \citep{Lightkurve2018}, asfgrid \citep{Sharma2016asfgrid, Stello2022b}, matplotlib \citep{matplotlib}, pyMON \citep{Howell2025}, TESSCut \citep{TESSCut}, echelle \citep{echelle1, echelle2}, isochrones \citep{Isochrones}
}

\appendix
\section{Inferring Full Seismic Masses}\label{sec:appendix}
\subsection{Measuring \dnu}\label{sec: Measuring dnu}
\begin{figure*}[hbt!]
\includegraphics[width=1.0\linewidth]{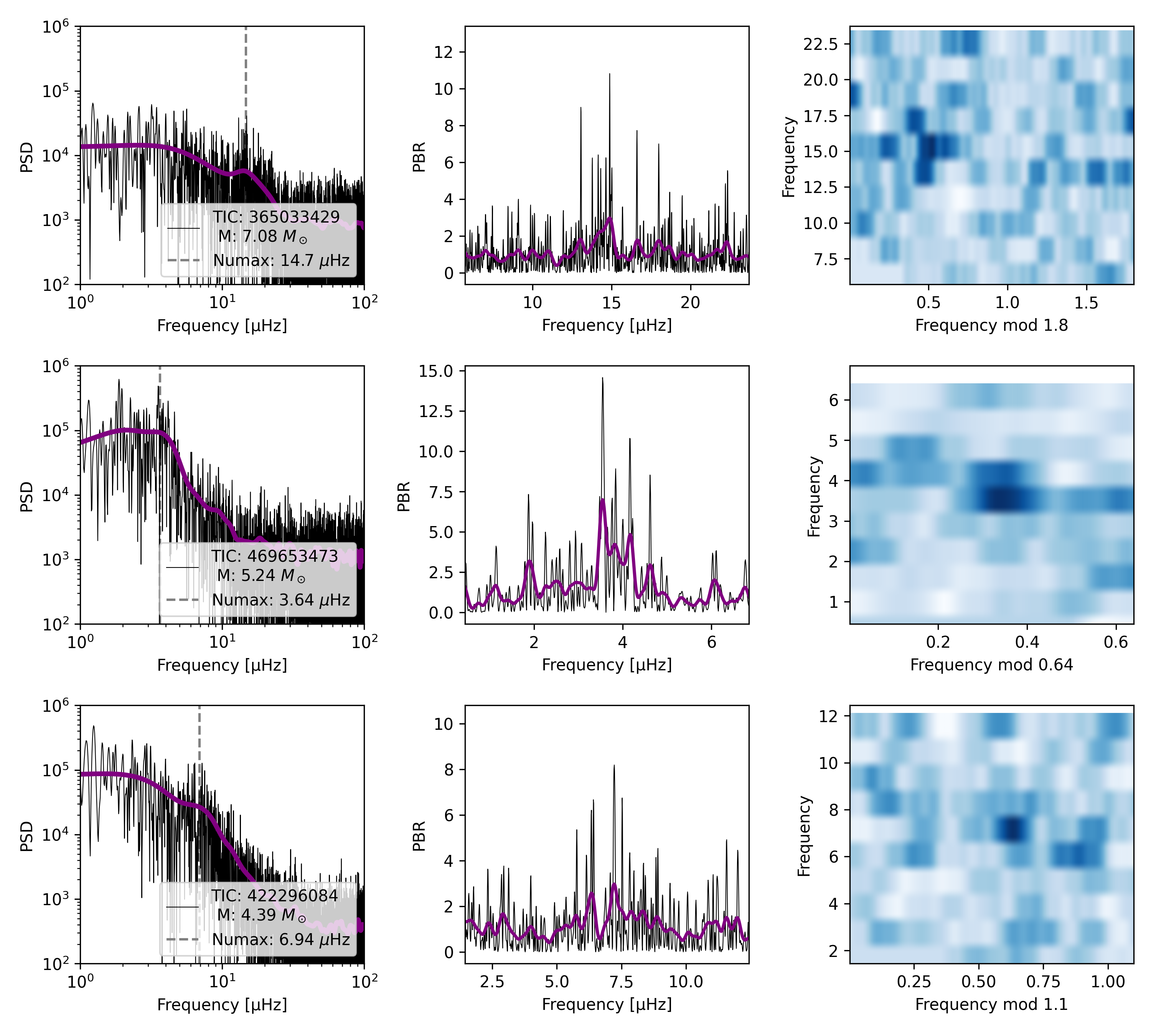}
\caption{Example power spectra (left), background-corrected power spectra (middle), and échelle diagrams (right) for three of the intermediate-mass stars in our sample. Power spectra are plotted in black and smoothed power spectra are plotted in purple. In log-log plots the power spectrum is smoothed with a gaussian filter of width equal to \dnu\ and in the background-corrected plots the power spectrum is smoothed with width equal to one-tenth \dnu. Listed masses are from equation \ref{eq:mass_ssr}. Échelle diagrams are generated using the \texttt{echelle} python package \citep{echelle1, echelle2} with our estimated \dnu.}
\label{fig:example_psd.png}
\end{figure*}

For \dnu, we compute the power spectrum of the power spectrum ($\rm{PS \otimes PS}$) around the oscillation envelope \citet{Hekker2010a, Mathur2010}, which we broadly define as $\pm5\ \Delta\nu_{\rm{est}}$ about \numax, where $\Delta\nu_{\rm{est}}$ comes from the \citet{Stello2009} \numax-\dnu\ relationship. We then choose the peak closest to $\Delta\nu_{\rm{est}}$ as an initial guess of the star's true \dnu. In cases where there was not a prominent peak, we adopt $\Delta\nu_{\rm{est}}$ as the initial guess. To verify our initial guess, we visually inspect the échelle diagram for each of our targets and then adjust \dnu\ accordingly until we begin to see ridges. We show a few example échelle diagrams in Figure \ref{fig:example_psd.png}. Since we measure \dnu\ by eye, it is difficult to estimate the uncertainty using bootstrapping methods like we did for \numax. Instead, we compute our \dnu\ uncertainties by scaling the median APOKASC-3 \citep{Pinsonneault2025} \dnu\ uncertainty ($\sigma_{\Delta\nu,\rm{Kepler}} = 0.6\%$) with the following equation:
\begin{equation}\label{eq:e_dnu}
    \sigma_{\Delta\nu,\rm{TESS}} = \sigma_{\Delta\nu,\rm{Kepler}}\frac{T_{\rm{Kepler}}}{T_{\rm{TESS}}}
\end{equation}
where $T_{\rm{Kepler}}$ is the four year \textit{Kepler} baseline and $T_{\rm{TESS}}$ is the baseline of each target star. 

We determine $f_{\Delta\nu}$ by entering metallicity, \teff, \numax, and \dnu\ into \texttt{Asfgrid} \citep{Sharma2016asfgrid, Stello2022b}, which determines the $f_{\Delta\nu}$ by interpolating across a grid of stellar models. However, because \texttt{Asfgrid} only covers uncorrected seismic masses (from the \teff, \numax, and \dnu\ inputs) up to $5.5\ \rm{M}_\odot$, 7 of our stars fall outside the model grid and are interpolated using the nearest grid point. These stars are flagged accordingly and we treat their mass estimates with caution. We then determine the full seismic radii and masses using equations \ref{eq:rad_dsr} and \ref{eq:mass_dsr}.

\subsection{Discussion on Full Seismic Masses}
\begin{figure}[hbt!]
\includegraphics[width=1.0\linewidth]{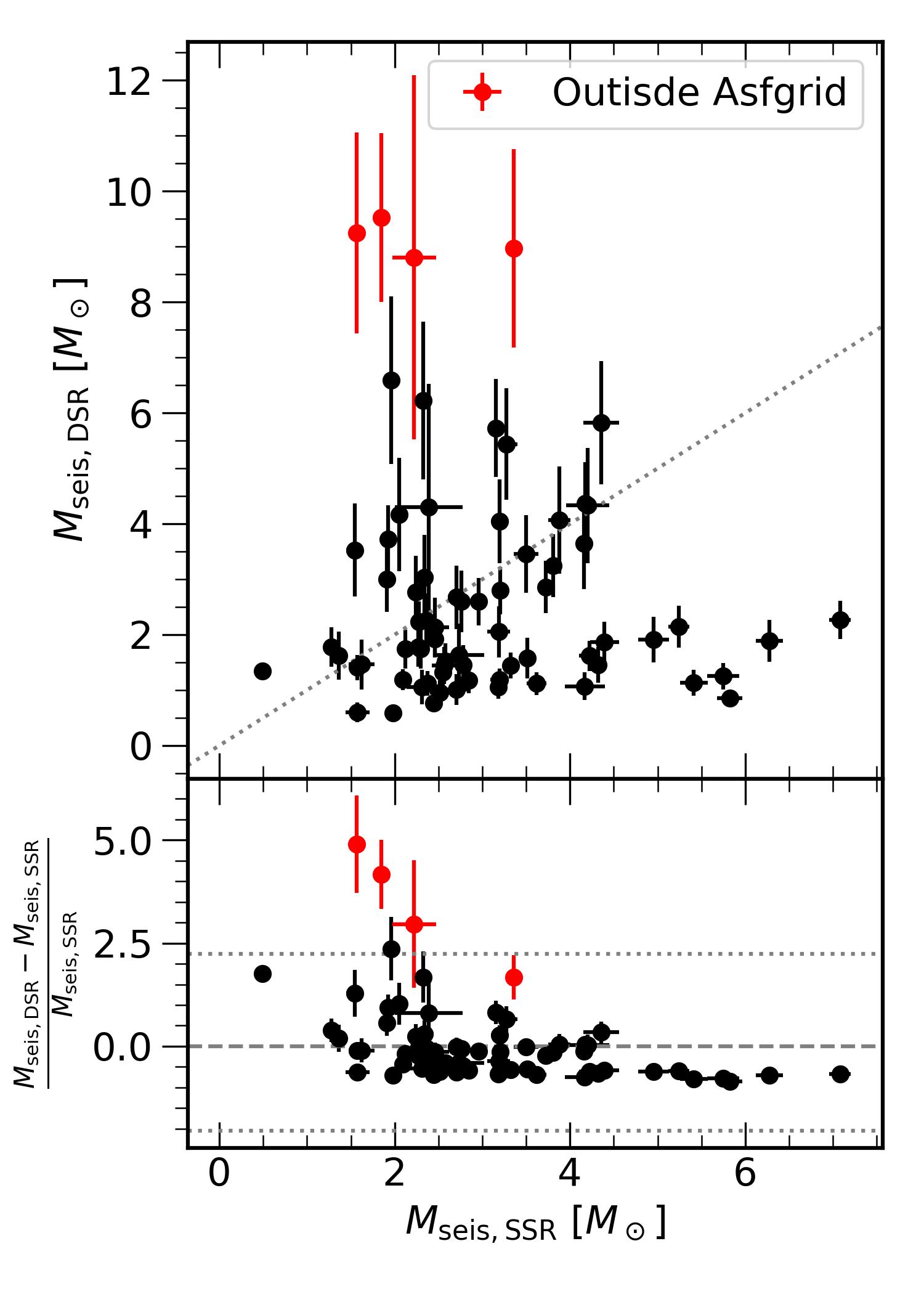}
\caption{\textbf{Top Panel:} Comparison of the DSR mass to the SSR mass for all stars with measured \dnu. Dotted line represents a 1:1 mass relation. \textbf{Bottom Panel:} Fractional mass residuals for all stars with measured \dnu. The dashed line represents the 1:1 mass relation and the two dotted lines correspond to $\pm2\sigma$ around the mean of the residual distribution. The mean fractional residual is $\sim0.10$. In both panels, the red points are the stars which have uncorrected seismic masses outside of \texttt{Asfgrid} as discussed in Section \ref{sec: Measuring dnu}.}
\label{fig:full_masses.png}
\end{figure}

Using the methods of Section \ref{sec: Measuring dnu} yields 71 stars with DSR mass estimates out of the 98 stars total with \numax\ detections. We compare the DSR mass estimates to the SSR mass estimates from \ref{sec: Asteroseismic Analysis} in Figure \ref{fig:full_masses.png}. It can be seen that there are significant deviations between the two mass estimates, which could be caused by a number of factors. 

The immediate one is that the \dnu\ measurements are not reliable. Since the bulk of our sample oscillates at very low frequencies, many of the \dnu\ measurements were determined by only two or three radial orders. This can be seen in Figure \ref{fig:example_psd.png}, where the ridges are not easily identifiable if they can even be identified at all.

Another potential source for systematic errors comes from the large radii that we see in these stars, with most of our sample falling between $20-80 \rm{R}_\odot$ for both the seismic and Gaia radii. The DSR begins to break down at large radii \citep{Zinn2023, Ash2025}, largely due to the \dnu\ term. In particular this causes the inflation of seismic radii leading to higher-mass estimates for stars with low \numax. 

As discussed in Section \ref{sec: Asteroseismic Analysis}, we assume $f_{\nu_{\rm{max}}}$ is of order unity. This could cause deviations, because the SSR depends on $f_{\nu_{\rm{max}}}$ linearly and the DSR depends on $f_{\nu_{\rm{max}}}$ cubed, the same correction factor will yield different seismic masses. 

It is not clear which of these effects could be the primary cause of the deviations in seismic mass estimates. As such, we choose to trust the SSR mass estimates over the DSR mass estimates.

\bibliography{bibliography.bib}
\bibliographystyle{aasjournalv7}

\end{document}